%% file: 0_main.tex
\documentclass[twocolumn,secnumarabic,amssymb, nobibnotes, aps, prd, superscriptaddress]{revtex4-1}
\usepackage{graphicx}
\usepackage{epstopdf}
\usepackage{amsmath}
\usepackage{amssymb}
\usepackage{xcolor}
\usepackage{soul}
\usepackage{ulem}
\usepackage{chngcntr}
\usepackage{appendix}

\begin{document}
\title{Effect of two parallel intruders on net work during granular penetrations}


\author{Swapnil Pravin}
\email{swapnil.pravin@temple.edu}
\affiliation{Temple University, Philadelphia, PA 19122}

\author{Brian Chang}
\affiliation{Temple University, Philadelphia, PA 19122}

\author{Endao Han}
\thanks{current affiliation: Joseph Henry Laboratories of Physics, Princeton University, Princeton, NJ 08544}
\affiliation{James Franck Institute, The University of Chicago, Chicago, IL 60637}

\author{Lionel London}
\affiliation{Massachusetts Institute of Technology, Cambridge, MA 02139}

\author{Daniel I. Goldman}
\affiliation{Georgia Institute of Technology, Atlanta, GA 30332}

\author{Heinrich M. Jaeger}
\affiliation{James Franck Institute, The University of Chicago, Chicago, IL 60637}

\author{S. Tonia Hsieh}
\email{tonia.hsieh@temple.edu}
\affiliation{Temple University, Philadelphia, PA 19122}

\begin{abstract}
\input{0_abstract}

\end{abstract}

\date{\today}
\maketitle

\begin{figure*}[ht]
\begin{center}
\includegraphics[width=1\linewidth]{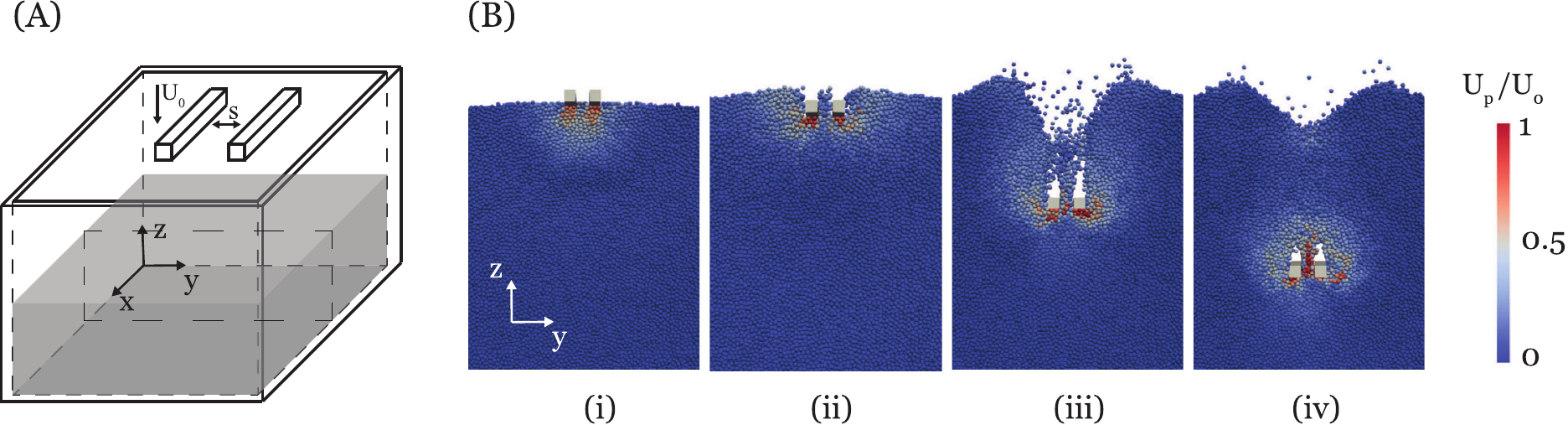}
\caption{(A) Schematic for the numerical simulation of two parallel rectangular prisms intruding into the surface of granular particles. The container has a horizontal cross-section of 25 cm x 25 cm, and is filled with spherical granular particles to a height of 10 cm. The intruders are moved vertically downward at a constant speed $U_0$=1 m/s. (B) A cross-section through the dashed box in panel A shows the granular particles colored by their instantaneous speeds $U_p$, normalized by the intruder speed $U_0$ when the intruder is at depths of $z=$ (i) 0.3 cm, (ii) 1.05 cm, (iii) 5 cm, and (iv) 8 cm. The total force on the two intruders was quantified from these simulations.}
\label{fig:schematics}
\end{center}
\end{figure*}

\input{1_intro}

\input{2_methods}
\input{3_results}

\input{4_conclusions}

\input{5_acknowledgments}

\clearpage
\appendix
\counterwithin{figure}{section}
\input{5_appendixA}

\clearpage
\input{5_appendixB}

\clearpage
\bibliographystyle{unsrt}
\bibliography{0_references}
\end{document}

%% file: 0_abstract.tex

The impact of single passive intruders into granular particles has been studied in detail. However, the impact force produced by multiple intruders separated at a distance from one another, and hence the effect of their presence in close proximity to one another, is largely unexplored. Here, we use numerical simulations and laboratory experiments to study the force response of two parallel rods intruding vertically into granular media while varying the gap spacing between them. We also explored the effect of variations in friction, intruder size, and particle size on the force response. The net work ($W$) of the two rods over the depth of intrusion was measured, and the instantaneous velocities of particles over the duration of intrusion were calculated by simulations. We found that the work done by the intruders changes with distance between them. We observed a peak in $W$ at a gap spacing of $\sim$3 particle diameters, which was up to 25\% greater than  $W$ at large separation (\textgreater 11 particle diameters), beyond which the net work plateaued. This peak was likely due to less particle flow between intruders as we found a larger number of strong forces---identified as force chains---in the particle domain at gaps surrounding the peak force. 
Although higher friction caused greater force generation during intrusion, the gap spacing between the intruders at which the peak work was generated remained unchanged. Larger intruder sizes resulted in greater net work with the peak in $W$ occurring at slightly larger intruder separations. Taken together, our results show that peak work done by two parallel intruders remained within a narrow range, remaining robust to most other tested parameters.

%% file: 1_intro.tex
The impact of a solid intruder into particulate media exposes the dual nature of granular media, that it can display characteristics of both solids and fluids during the process of intrusion \textcolor{blue}{\cite{vandermeer2017}}. An intruder passively falling into a granular bed under gravity experiences a strong drag force which brings the intruder to rest \cite{bester2017collisional,Clark2014,Deboeuf2009,Goldman2008,Katsuragi2007,pacheco2010cooperative,Royer2011,newhall2003projectile,ambroso2005dynamics,tsimring2005modeling,seguin2008influence,brzinski2013depth,tiwari2014drag,vandermeer2017}. For active intrusion under constant velocity, the force-depth relationship beyond a brief transient associated with the initial impact, is typically found to be linear and independent of velocity, even for intrusion speeds well beyond the quasi-static regime \cite{Roth2019,kang2018archimedes,aguilar2016robophysical}. The vast majority of these studies are focused on a single intruder. On the other hand, the force response to multiple intruders separated by a distance is poorly understood. Some previous works that have explored multiple intruders indicate the presence of attractive and repulsive forces between intruding disks \cite{LopezDeLaCruz2016}, spheres \cite{dhiman2020origin}, and a sphere and a wall \cite{Nelson2008}. Additional studies demonstrate a characteristic length scale at which two intruders begin to interact with one another during intrusions into bidimensional granular packing \cite{Merceron2018}.

Active intrusion of impacting solids into granular media has direct relevance for the terradynamics of animals as well as for development of robotic locomotors \cite{li2013terradynamics,aguilar2016robophysical,li2010effect}. In biological systems, interactions between multiple intruders are more common than intrusions by single, simple geometries. For example, feet often have toes which act as multiple intruders upon ground impact with each step. There is an enormous diversity of foot and toe morphologies in the animal kingdom, and toes likely serve an important function in the mechanics of interaction of feet with granular media \cite{Li2012}. In addition to contributing towards elucidating evolutionary drivers of biomechanical and morphological diversity, understanding the physics of the interactions of toes with granular media during a step has important implications for the design of robotic feet.

In this paper, we study the drag force on two co-intruding objects separated by a certain distance. We performed numerical simulations and experiments for two parallel rods actively intruding into dry granular media. We expect a non-monotonic dependence of the drag force on the distance between the two intruders because of the competition between two effects: increasing the intruder spacing from zero increases the effective cross-sectional area if the particles between the intruders remain hindered in their movement, but the likelihood with which that can happen decreases with intruder spacing. Therefore, one may expect a peak in force at some intruder spacing. This non-monotonicity of the drag force, and the location of its peak, has not been explored in detail before and is the focus of this paper. Additionally, we examine how the force response is influenced by intruder shape, intruder size relative to particle size, and inter-particle friction within the granular medium.In these experiments and simulations the particle size was chosen sufficiently large that the role of the interstitial air could be neglected.

%% file: 2_methods.tex
\section{Methods}

\subsection{Numerical simulations}
The 3D discrete element method (DEM) open-source software package LIGGGHTS\textsuperscript{\textregistered} was used to simulate the movement of particles. First, the granular bed was prepared by randomly generating spherical particles with a diameter of $d_g$=2 mm to above a container and allowing them to fall and settle under gravity. Actual poppy seeds have a more complex shape than a sphere. The particle parameters used in the simulations are listed in Table \ref{tab:sim_parameters}. Once the kinetic energy of the particles in the container decreased to nearly zero, two parallel rods ($D_r$ = 5 mm, $L_r=$ 5 cm), placed at a distance of $s$ apart, vertically intrude into the granular bed at a constant speed of $U_0$=1 m/s to a depth of $z_f$=8 cm (Fig  \ref{fig:schematics}A). 
 
 The force between two granular particles $i$ and $j$ is calculated as the sum of normal and tangential forces.


\begin{equation}
 \vec{F}_{ij} = (k_n \delta n_{ij} - \gamma_n v_{n,ij})\hat{n} + (k_t \delta t_{ij} - \gamma_t v_{t,ij})\hat{t}
\label{eqn:force_model}
\end{equation}

Each term within the parentheses contains a spring force and a damping force. $k_n$ and $k_t$ are the elastic constants for normal and tangential contacts, respectively. $\gamma_n$ and $\gamma_t$ are the viscoelastic damping constants for normal and tangential contacts. $\delta n_{ij}$ is the normal overlap of the two particles. $\delta t_{ij}$ represents the tangential displacement between the particles for the duration they are in contact, and is truncated to satisfy $F_t \leq \mu F_n$, where $F_t$ and $F_n$ are the tangential and normal forces respectively, and $\mu$ is the friction coefficient. A Hertzian contact force model is represented by the terms $k_n\delta n_{ij}$ and $k_n\delta t_{ij}$, where $k_n,k_t \propto \sqrt{\delta n_{ij}}$ as described in equations B1 and B3 in Appendix B. Normal and tangential components of relative velocity between two particles are denoted by $v_{n,ij}$ and $v_{t,ij}$, respectively. $\hat{n}$ is the unit normal vector and $\hat{t}$ is the unit tangential vector. 

The coefficients $k_n$, $k_t$, $\gamma_n$, and $\gamma_t$ are calculated from the material properties as described in appendix B. The numerical time step used in the simulations was $dt=5 \times 10\textsuperscript{-6}$ s.




\begin{table}
\caption{Properties of the granular particles used for DEM simulations. Values in parenthesis used for parameter sweep.}
\label{tab:sim_parameters}
\begin{center}
\begin{tabular} {l l}
Property & Value \\
\hline
Rod length, $L_r$           & 5 cm \\
Rod diameter or width, $D$  & 5 mm (1-6 mm) \\
Particle diameter, $d$		& 2 mm (4, 6 mm) \\
Bulk density, $\rho$		& 1100 kg m\textsuperscript{-3} \\
Volume fraction, $\phi$     & 0.62 \\
Young's modulus, $E$ 		& 5 x 10\textsuperscript{6} Pa\\
Poisson's ratio, $\nu$ 		& 0.3 \\
Coefficient of restitution 	& 0.2 \\
Coefficient of friction , $\mu$ 			& 0.5 (0.1-1) \\
Timestep, $dt$              & 5x10\textsuperscript{-6} s \\
Spacing (varies), $s$                & 0-20 \\

\hline
\end{tabular}
\end{center}
\end{table}

\begin{figure}[ht]
\begin{center}
\includegraphics[width=1\linewidth]{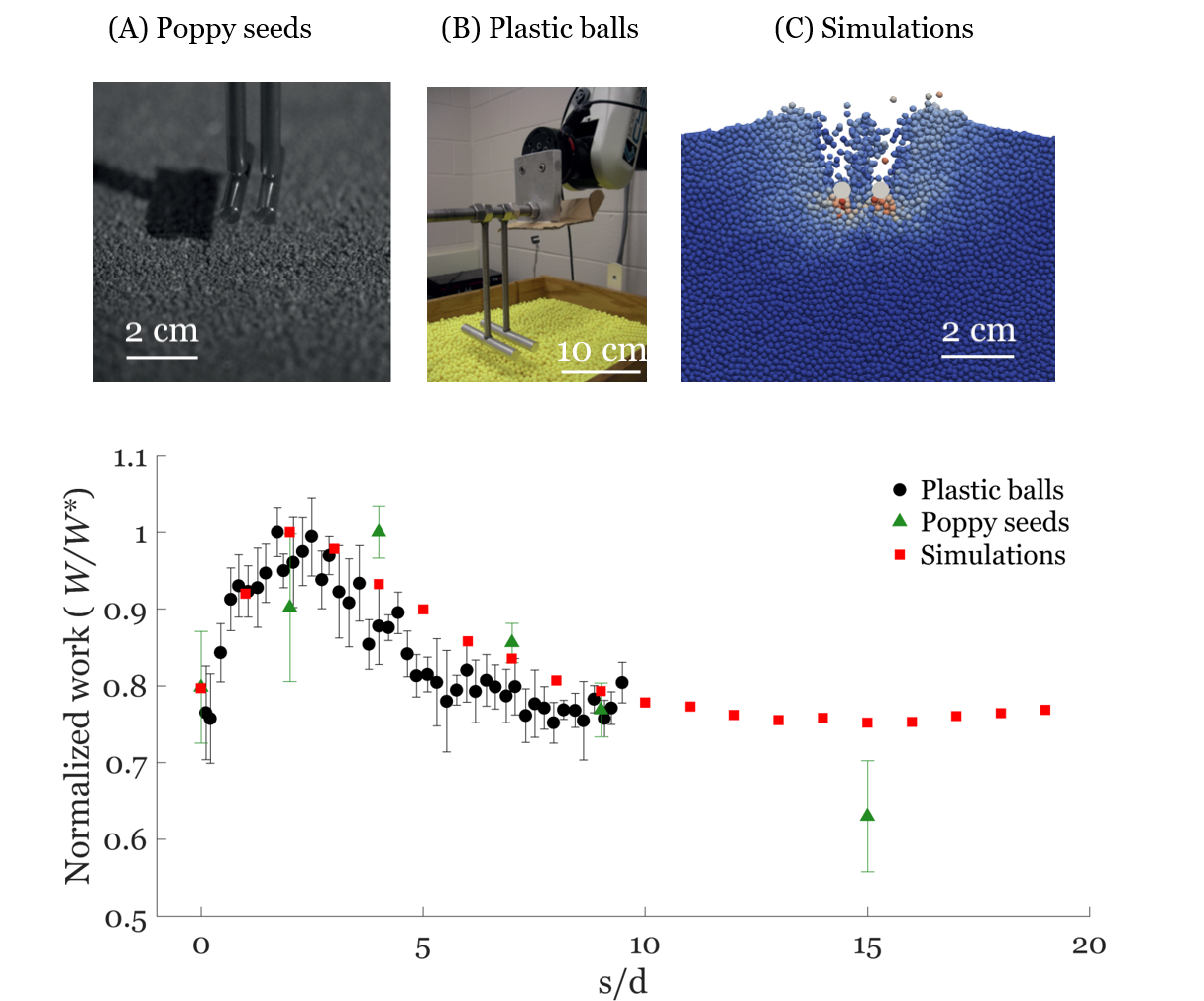}
\caption{
Comparison between simulation and experiment for net work ($W$) by the parallel rods during vertical intrusion for (A) poppy seeds ($d_{ps}\approx 1$ mm) at $U_0=1$ m/s, (B) plastic balls ($d_{pb}=6$ mm) at $U_0=0.18$ m/s, and (C) simulated spherical particles ($d_{g}=2$ mm) at $U_0=1$ m/s. Each curve is normalized by its peak value ($W^*$). $W^*$ is 1.05 J for poppy seeds, 3.32 J for plastic balls, and 0.55 J for numerical simulations. The gap between the rods is normalized by the particle diameter. For numerical simulations, the very strong forces are shown for s/d=3, 6, and 15 at a depth of 4 cm. A peak in the force response is observed in each case between 2 and 4 particle diameters.
}
\label{fig:expt}
\end{center}
\end{figure}

\subsection{Granular intrusion experiments}
To validate our simulation results, we performed two sets of experiments using parallel cylindrical rods vertically intruding at constant speed into a container of (a) poppy seeds at 1 m/s, and (b) plastic ball bearings at 18 cm/s.

\subsubsection{Intrusion into poppy seeds}
Poppy seeds with a diameter of 0.8-1.6 mm were poured into the container and then the container was shaken sideways with an exponentially decaying amplitude to relax the sample and obtain a flat top surface. The overall volume fraction of the sample was 0.62. Two circular cross-section aluminum rods of 0.5 cm diameter and 3 cm length were used as intruders. The intruders were mounted to a linear actuator (ETT050, Parker Hannifin Corp., Cleveland, OH) and moved vertically downward at a constant speed of 1 m/s. A  force  transducer (DLC101-100, Omega Engineering, Inc., Norwalk, CT) was used to measure the instantaneous force on the intruders for the duration of impact. The granular media had a depth of 13 cm, and the intruders were pushed to a depth of 8 cm from the top surface. Force measurements were made for gap spacings of $s/d$= 0, 2, 4, 7, 9, and 15.

\subsubsection{Intrusion into plastic balls}
A container was filled with 6 mm diameter plastic spheres. Two circular cross-section aluminum rods of 2.54 cm diameter and 9.65 cm length were rigidly mounted to a robotic arm (CRS Robotics, Ontario, Canada). The robotic arm moved vertically downward at a constant speed of 18 cm/s with an intrusion depth of 10 cm through the plastic balls. The force response (ATI Industrial Automation, Apex, NC) at various intruder separations, $s$ was recorded.  

%% file: 3_results.tex
\section{Dependence of work on spacing}
To study the effect of intruder gap on force response, we performed multiple simulations of intruders separated at different gap spacings, and examined the dependence on intruder shape, size, and particle friction. The net work ($W$) by the intruders over the depth of intrusion was calculated as $W=\int_0^{z_f}F(z) \, dz$, where $F(z)$ is the instantaneous force experienced by the intruders, $z$ is the vertical distance from the surface of granular substrate, and $z_f$=8 cm is the fixed depth of intrusion throughout all simulations. The net work, $W$ for each gap spacing for cylindrical intruders is shown in figure \ref{fig:expt}. There is good agreement between the simulation and the experiments of two cylindrical intruders intruding into a bed of spherical particles, despite the differences in intrusion speed and particle size.
We find that the maximum $W$ occurs around $s/d\sim2$ for the simulation and experiments on spherical particles, while the intrusion experiments on poppy seeds have a maximum work around $s/d\sim4$. Previous works show similar trends in maximum force production, but in different systems. For example, in an earlier work on two spheres intruding into a granular bed, a maximum attraction force was experienced at a separation of approximately 4 particle diameters\cite{dhiman2020origin}. The attraction force subsequently decreased as the separation distance increased. While this phenomenon appears robust in different systems, we use simulations to further explored how various other particle and intruder configurations could affect the non-monotonic relationship between force and intruder separation distance.

\subsection{Intruder shape}
Intruder shape is known to influence intrusion dynamics. For example, when a conical intruder impacts a granular surface, as the slope of the intruder tip relative to the granular media surface increases, a smaller drag and a deeper penetration is observed \cite{bester2017collisional}. Another study using a hemispherical disc and photoelastic particles showed large force fluctuations emanating from the leading edge of the intruder in directions dependent on the local slope of the intruder edge \cite{Clark2012}.

\begin{figure}[ht!]
\begin{center}
\includegraphics[width=.8\linewidth, trim={0 8cm 19cm 0}]{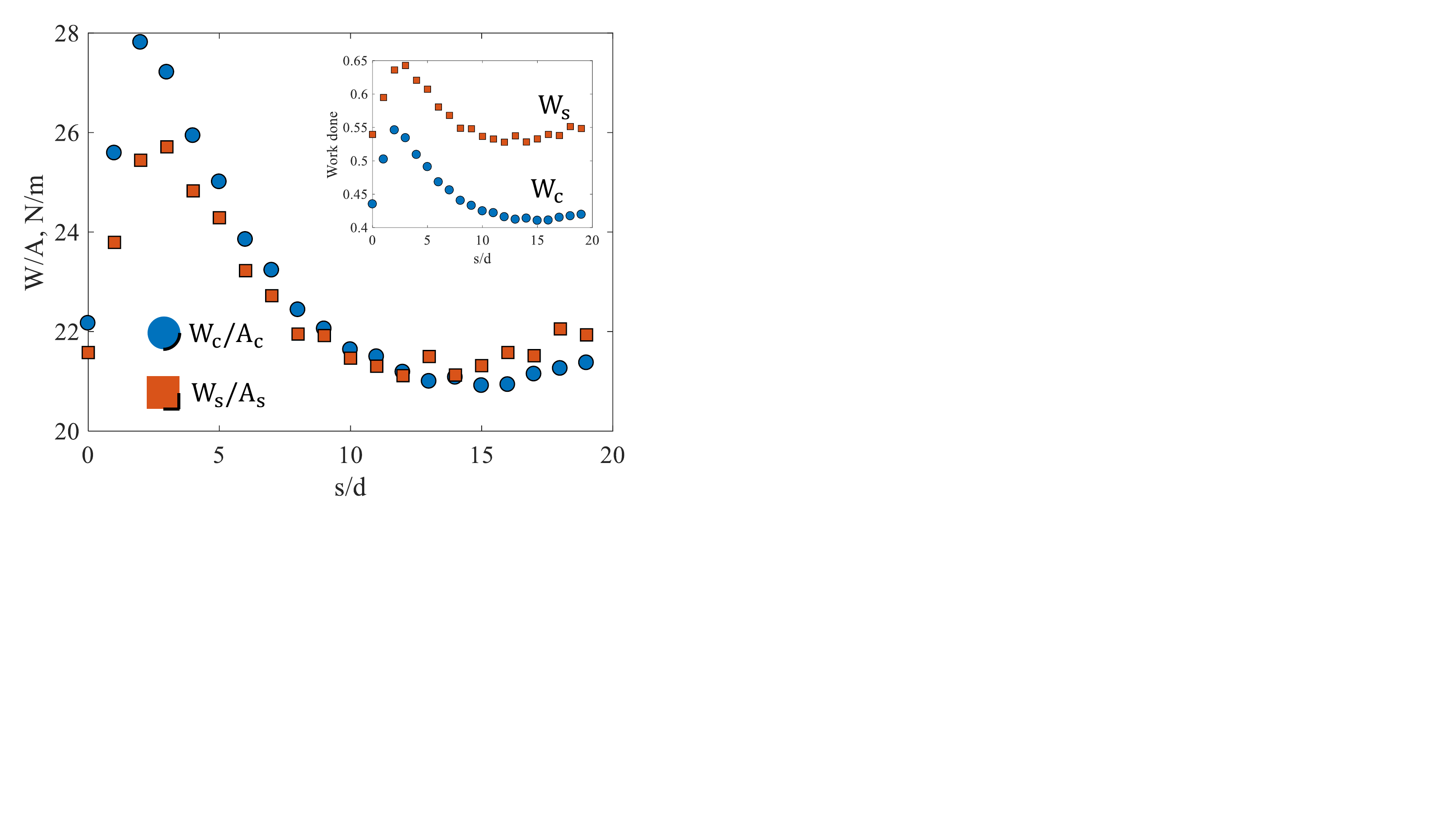}
\caption{Net work ($W$) over the depth of intrusion for the square ($W_s$) and cylindrical ($W_c$) shaped intruders in simulations. The ratio of the surface areas of the two shapes is $4/\pi$. Scaling the work done by the cylindrical shape with this factor nearly collapses the two curves on one another. Inset: While both shapes display a peak in force at $\sim$3 particle diameters, the square rods experience a greater overall force. The friction coefficient was 0.5 for both shapes. }
\label{fig:shape}
\end{center}
\end{figure}

To determine how intruder shape influences multi-body intrusion dynamics, we compared two basic shapes: square and cylindrical rods. This choice was largely driven by the consideration that the square shape would produce force chains anchored to its bottom surface and therefore the two sets of force chains would be largely parallel to one another and interact minimally. The cylindrical shape on the other hand would produce force chains in the sideways direction as well, emanating at angles relative to $z$, thus leading to greater ``interaction'' among the two sets of forces. 

Figure \ref{fig:shape} shows that the work done by both the geometrical shapes has a peak near three particle diameters of intruder gap. Although the general behavior of the curves is similar, the square rods experience a greater force for all gap spacings, as shown in the inset of figure \ref{fig:shape}.

It is reasonable to expect that the forces generated by the two shapes would be proportional to the respective surface areas on the two intruder shapes where the force chains originate. Figure \ref{fig:shape} shows the areas of interest where the force chains would be expected to originate. Following this assumption, the surface area of the square and circle intruders would be $A_s=2R \,L_r$ and $A_c=\pi/2 R L_r$. 
By dividing work by the corresponding surface areas, we find that the two curves collapse quite well when $s/d>5$ and when $s/d = 0$ (figure \ref{fig:shape}), indicating that the average pressure is independent of geometry when the intruders can be treated as independent ($s/d \gg 1$) or be treated as one ($s/d < 1$). 
In between, we note that circular intruders produce more work per area than the square ones. 
In this intermediate regime, the effective area of the intruders is increased because the higher resistance the grains experience when they are squeezed through the gap in between the two rods. 
The difference in $W/A$ indicates that cylindrical rods generate either denser force chains in between (see Sec.~\ref{sec:ForceChains}), thus creates a larger relative ``effective area''.

\begin{figure}[ht!]
\begin{center}
\includegraphics[width=.8\linewidth, trim={0 1cm 17cm 0}]{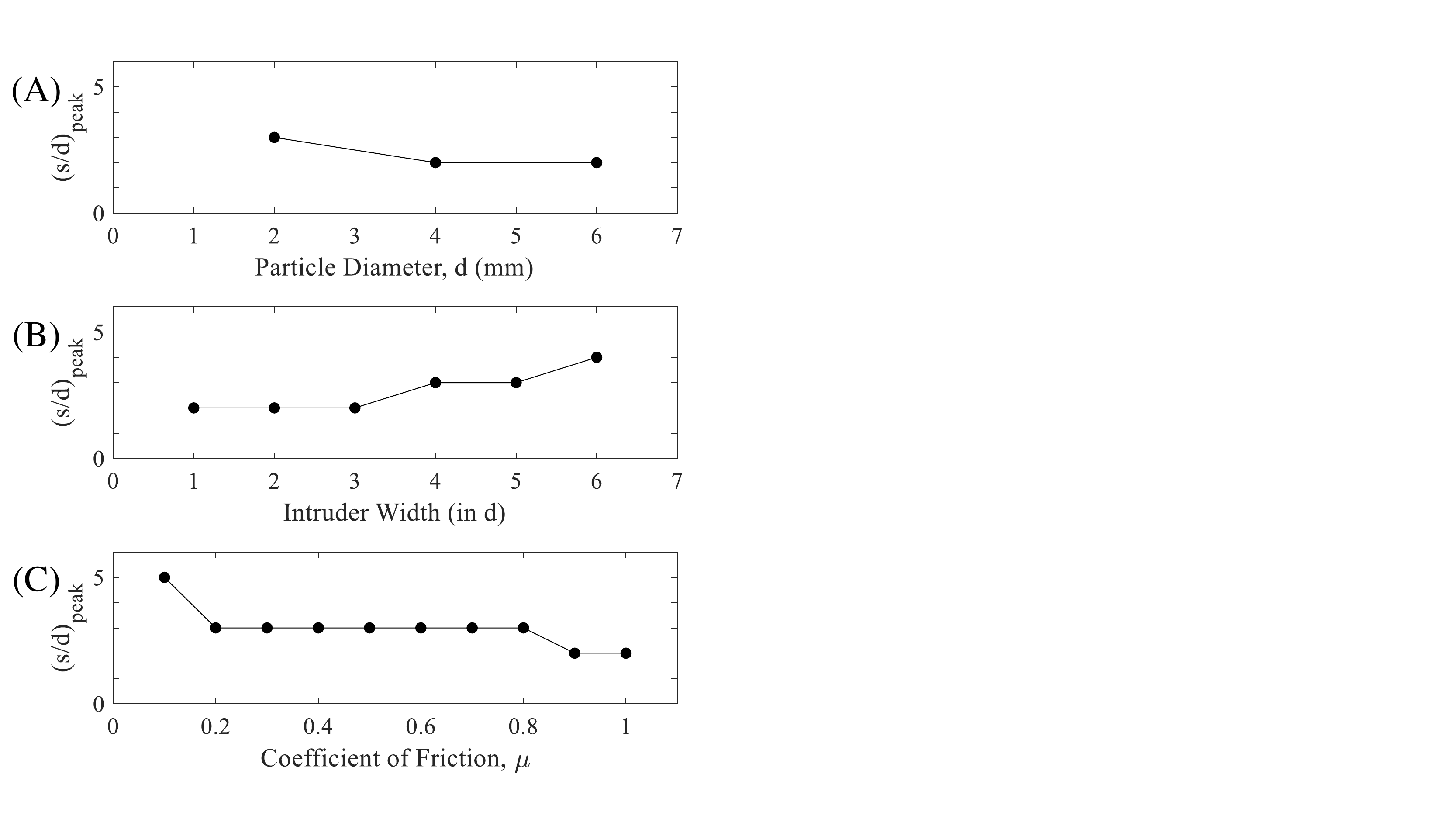}
\caption{Spacing ($(s/d)_{peak}$) that corresponds to the peak force observed with respect to (A) particle diameter, (B) intruder width, and (C) coefficient of friction from numerical simulations. Each respective work vs spacing plot is shown in Appendix A.1-A.3.  }
\label{fig:sweep}
\end{center}
\end{figure}


\subsection{Intruder and particle size}
The effect of particle size ($d=$ 2, 4, and 6 mm) on the force response was explored with simulations while keeping the particle density and intruder size constant. The size of the particle domain was expanded for larger particle sizes to avoid wall effects. We find that the magnitude of work increases with particle size (Appendix figure \ref{fig:SI_particled}), which may be a consequence of increasing particle mass. A previous study showed that particle diameter has little effect on the force of intrusions within the quasi-steady regime, $v<\sqrt{2 g \,d}/10$ \cite{feng2019support}. While particle density can have a significant affect on the force-depth profile, curves collapse well with a normalizing factor that includes density.

The particle sizes of 4 and 6 mm are roughly the same size of the intruder width (5 mm) in these simulations. We chose to simulate these particular particle sizes because when the particle size is larger than the intruder size, we find that there are multiple force peaks not observed empirically (as seen in Appendix \ref{fig:SI_particled}). We define $(s/d)_{peak}$ as the spacing at which peak work occurs. By examining the first peak, we find that $(s/d)_{peak}$ remains relatively constant over a factor of 3 change in particle diameter (figure \ref{fig:sweep}A).

To examine the effect of intruder size, the horizontal width, $D$, of the intruder was changed while keeping other parameters constant. The net work, $W$ increases with increasing intruder size (Appendix figure \ref{fig:SI_intruderD}). While the nature of the curves is preserved at higher intruder sizes, $(s/d)_{peak}$ increases with intruder widths exceeding 3$d$ as seen in figure \ref{fig:sweep}B.

\subsection{Friction}
To examine the role of friction in resistance to intrusion, we performed the granular impact simulations with different particle friction coefficients, $\mu$. Both intruder friction \cite{zheng2018intruder} and particle friction coefficient \cite{li2013terradynamics,clark2016steady} have been shown to affect the formation of force chains originating from the intruder surface. We hypothesized that the spacing at which a peak in net work ($s/d_{peak}$) occurs would increase with increasing friction coefficient, as the particles would form longer force chains with increasing $\mu$. Interestingly, we found that the gap spacing at which the peak work occurs changes appreciably only at the very low and very high friction coefficients (figure \ref{fig:sweep}C). 
The probability distribution of inter-particle forces in figure \ref{fig:SI_PDF} confirms that the force magnitudes increase with greater friction. This leads us to conclude that force chains are stronger for higher $\mu$, but their length does not increase appreciably with increasing $\mu$.  

\section{Velocity profiles}

\begin{figure}[ht]
\begin{center}
    \includegraphics[width=.3\textwidth, trim={3cm 0cm 20cm 0}]{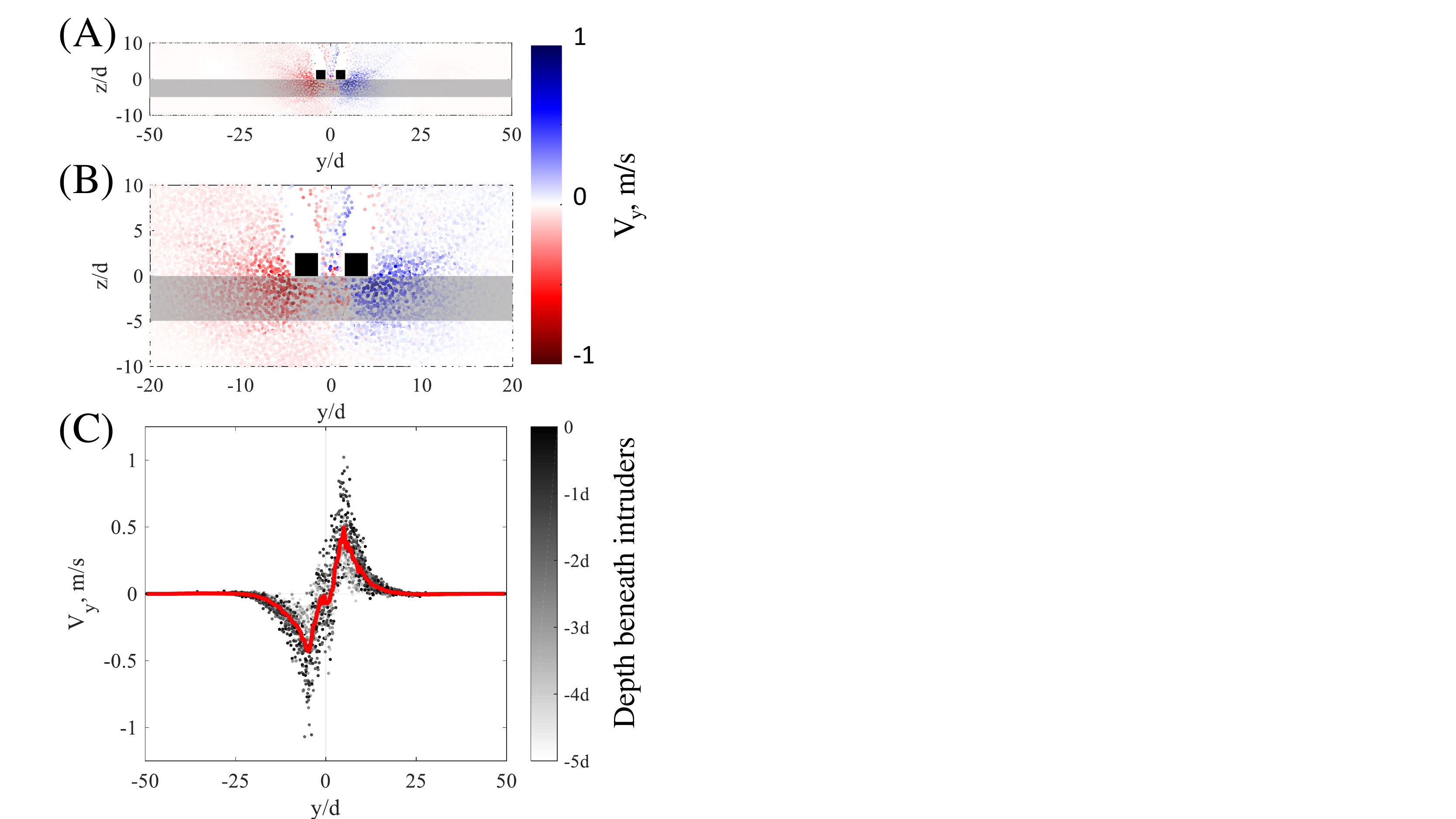}
    \caption{Velocity flow fields from simulations. (A,B) Y-direction velocity, $V_y$, of particles within a region of $x/d=-10$ to $10$ at a spacing of $s/d=3$ and depth of $z=4$ cm. (C) The velocity profile of the particles within the shaded region below the intruder, which has a height of 5$d_g$. Gray-scale colorbar represents the depth of the particle relative to the intruder, such that black points are particles directly beneath the intruder and white points are particles $5d$ below the intruder. The red line is the average velocity of the particles.
    }
    \label{fig:particles_VyVz}
    \end{center}
\end{figure}

To gain more insight into the physical mechanisms causing the peak in work done around $s/d=3$, we examined the velocity profile of the particles directly below square intruders. Average y-direction particle velocities, $V_y$, within the region $x/d=[-10, 10]$ are shown in figure \ref{fig:particles_VyVz}. Particle velocities within $5d$ directly below the intruders are highlighted in gray in figure \ref{fig:particles_VyVz}A, and then plotted in figure \ref{fig:particles_VyVz}B. The velocity profile has little dependence on depth, as shown in figures \ref{fig:profile_vy} and \ref{fig:profile_vz} of appendix \ref{sec:appA}. One might expect a transient response such that the velocity magnitude grows and decays over time or depth. While there is some evidence of this at $z=1$ cm, the velocity profiles quickly approach a steady state behavior as the intruders go further into the substrate. Therefore, all analysis carried out will consider the moment at $z=4$ cm.

\begin{figure}[ht]
\begin{center}
    \includegraphics[width=.3\textwidth, trim={0 0 20cm 0}]{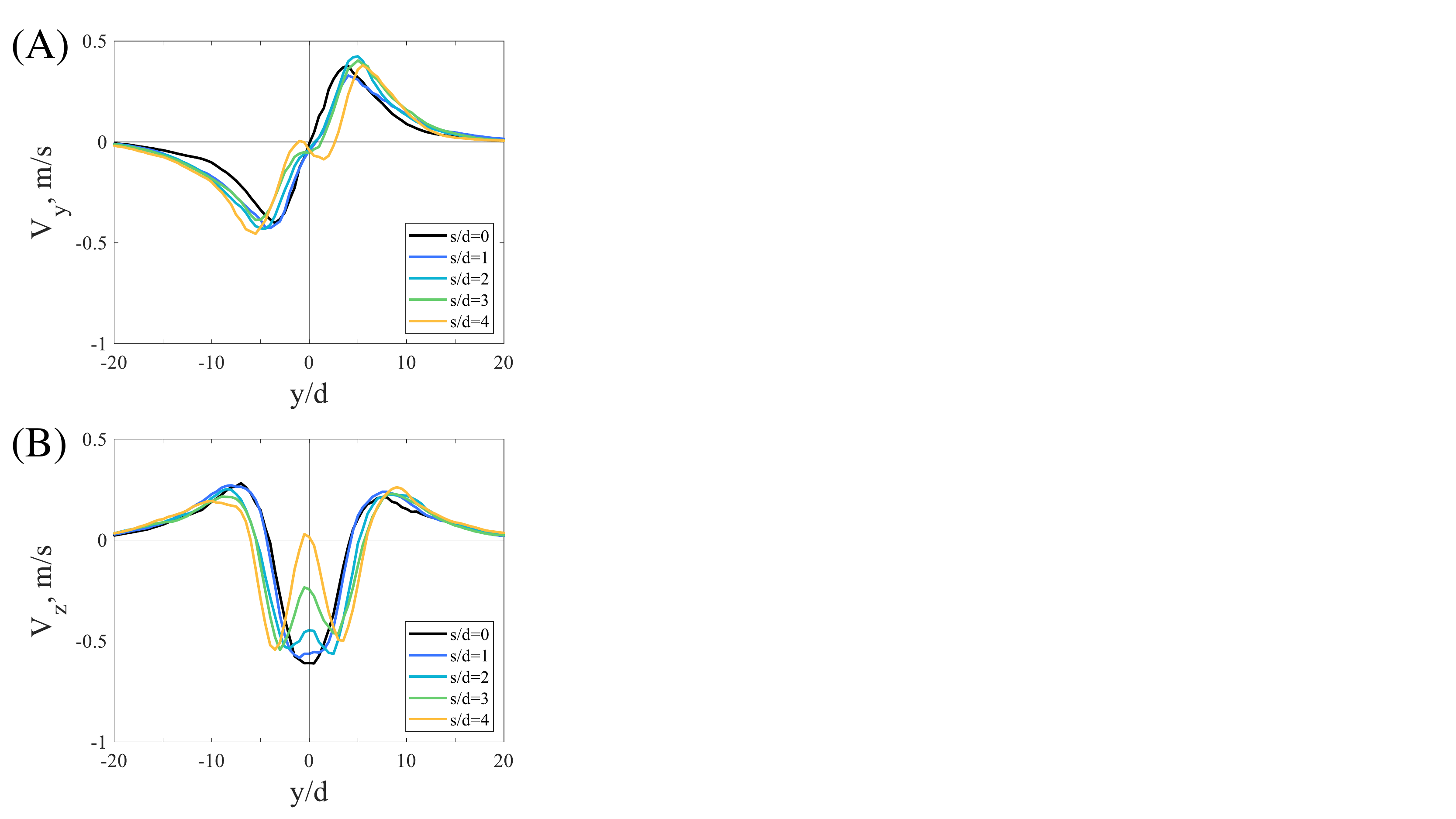}
    \caption{Average velocity profiles at instantaneous depth of $z=0.5 z_f$. (A) Y-direction velocity profile, $V_y$. Increasing $s/d$ begins to show a development of an inflection point between the two intruders. Near the center, $y/d=0$, the slope of $V_y$ transitions from positive to negative when $s/d>3$. (B) Z-direction velocity profile, $V_z$. Increasing $s/d$ causes $V_z$ between the intruders to begin changing directions relative to the direction of the intruder motion. Near the center, $y/d=0$, $V_z$ transitions from negative to positive when $s/d>3$.
    }
    \label{fig:v_trans}
    \end{center}
\end{figure}

\begin{figure*}[ht]
\begin{center}
\includegraphics[width=.95\textwidth]{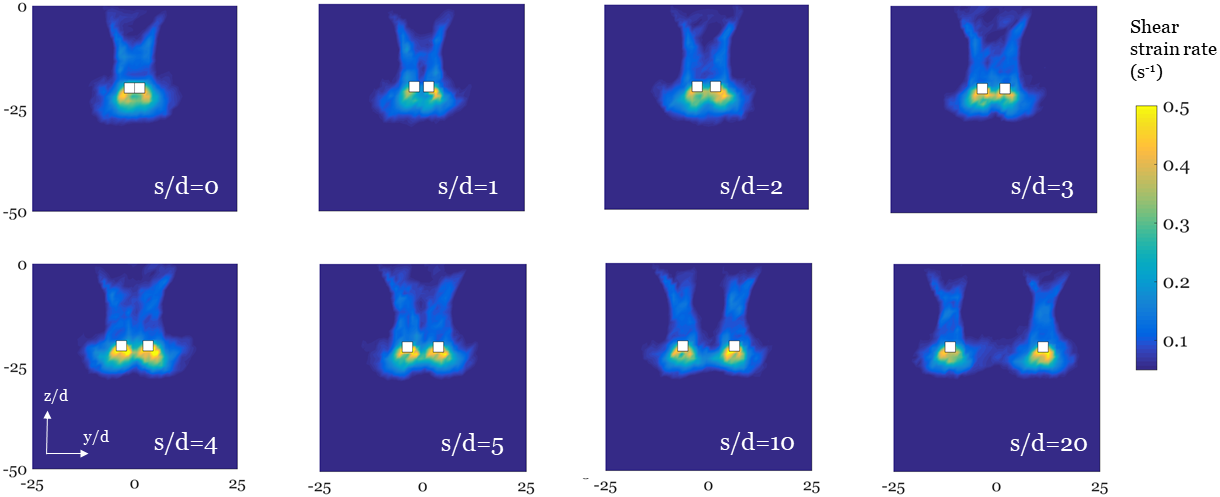}
\caption{Shear strain rate, averaged along the length of the intruders, at $z = 4$ cm for various intruder spacings ($s/d$). The y and z axes are normalized by particle diameter, $d$. 
}
\label{fig:strain}
\end{center}
\end{figure*}

\begin{figure}[ht]
\begin{center}
    \includegraphics[width=.8\linewidth]{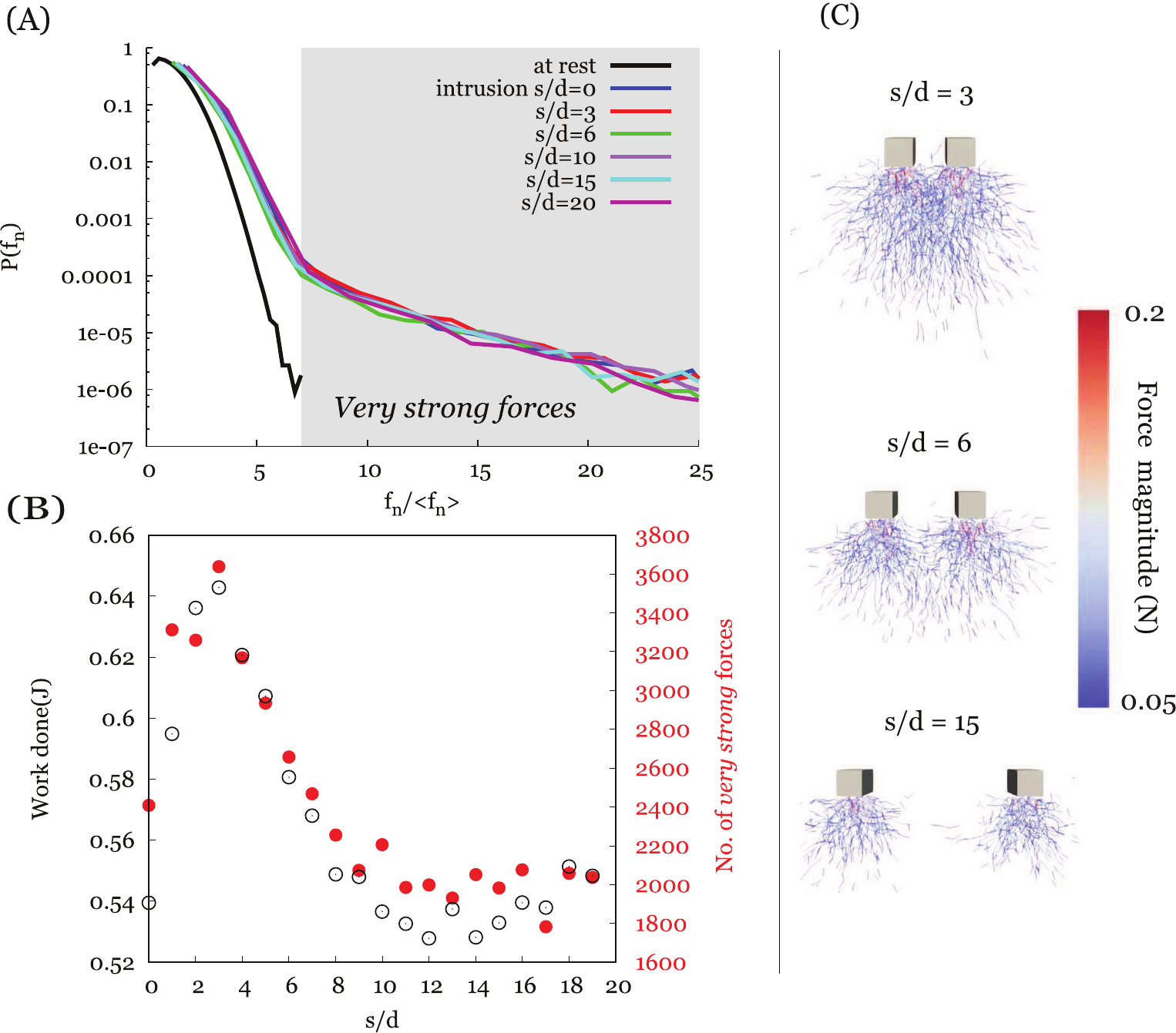}
    \caption{(A) The probability density function of normal forces, both for particles at rest, and those undergoing intrusion at intruder spacing of $s/d=$0,3,6,10,15 and 20, at the instant when the intruders are at a depth of 4 cm. In addition to the exponential decay of strong normal forces (mean force $>$ 1), a tail in the force distribution is observed during intrusion. These very-strong forces (larger than $\sim 6 <f_n>$) are caused by the active intrusion. (B) The left axis shows the net work of two square intruders. The right axis shows the number of normal forces between particles that form the tail of the force distribution (very-strong) for each intruder separation. The two curves follow a similar pattern, indicating that the very strong forces and the resulting smaller particle flow could be responsible for the peak in force observed around $s/d\sim3$. (C) The strong forces for $s/d=$3,6, and 15. }
    \label{fig:pdf}
\end{center}
\end{figure}

Figure \ref{fig:v_trans}A shows the average Y-velocity profile, $V_y$ at $0 \leq s/d \leq 4$. At $s/d=0$, when the intruders are adjacent and touching each other, the average particle velocity switches from negative to positive near the center ($y/d=0$), owing to the fact that the half of the particles move towards the left (-) and the others move toward the right (+). Particles also exhibit local minima and maxima, which are near the edges of the intruder. By increasing the spacing, we find that an inflection point begins to develop near $y/d=0$, such that the slope transitions from positive to negative when $s/d=4$ (figure \ref{fig:v_trans}A) and a new set of local minima and maxima appear. This indicates the critical spacing at which particles begin to flow toward the center instead of away from it.

A similar transition occurs for the $V_z$ velocity profile (figure \ref{fig:v_trans}B). At $s/d=0$, the velocity profile exhibits a minimum near $y/d=0$, directly below the intruders. Previous works have observed similar behaviors \cite{kang2018archimedes}. Traditionally, particles moving in the same direction as intruder motion at the same speed is a possible sign of jamming \cite{aguilar2016robophysical,kang2018archimedes}. At other spacings of $s/d=1$ to 3, $V_z$ was also negative between the intruders, indicating particles between the intruders were moving largely with the intruders but at a slower velocity, suggesting incomplete jamming. However, increasing the spacing causes particles to transition from negative to positive when $s/d>3$. This shows that on average, particles between the two intruders will move upwards, indicating particle flow between the intruders. This also correlated with the decrease in force at $s/d=$ 4.

\section{Shear strain rate}

The data generated by the simulation was re-sampled onto a structured volume grid to facilitate the calculation of derivatives throughout the particle domain. The 3D shear strain rate was calculated from the re-sampled velocity data as

\[  \bar {\bar{\epsilon} } = \frac{1}{2} (\nabla u + (\nabla u)^{T}) \]

where $\nabla u$ is the velocity gradient tensor. The magnitude of the strain rate tensor was calculated using the continuum mechanics definition of a tensor magnitude ($||A|| = \sqrt{A:A}$).

\[ |\bar{\bar{\epsilon}}| = \sqrt{ \epsilon_{ij} \epsilon_{ij} } = \sqrt { \epsilon_{11}^2  + \epsilon_{22}^2 + \epsilon_{33}^2 + 2\epsilon_{12}^2 + 2\epsilon_{23}^2 + 2\epsilon_{13}^2 } \]

Figure \ref{fig:strain} compares the average strain rates along the length of the intruders for $s/d=$ 0, 1, 2, 3, 4, 5, 10, and 20. 
When the gap size is less than the particle diameter ($s/d < 1$), no particle can pass between the square rods.
In this case, a stagnation zone \cite{kang2018archimedes,aguilar2016robophysical} is observed below the intruder where the shear rate is significantly smaller than the surrounding flow due to little relative motion between particles. 
This increases the effective area of the intruder while pushing the particles. 
As the gap size increases, particles are able to pass through, but stronger force chains can be built intermittently, as will be shown in Section~\ref{sec:ForceChains}, which leads to higher resistance to the granular flow. 
As a result, the effective area is still above the combined surface area of the two rods. 
When the two rods are more than 10 particle diameters apart, the interactions between the flows generated by individual rod are less significant, and they can be treated as independent intruders. \\

\section{Role of Strong Force Chains}
\label{sec:ForceChains}
We further investigated the possible role of strong forces that may lead to impeded particle flow between the two intruders by examining the probability density distribution of normal forces (Fig. \ref{fig:pdf}). Strong forces, which we define as normal forces greater than $<f_n>$, show an exponentially decreasing distribution, as observed in previous works \cite{radjai1999contact,radjai1998bimodal}.
We observed a set of forces following an inflection in the force distribution curve for which the normalized force distribution is not significantly different among the various intruder spacings, and typically occurs after 6$<f_n>$ (Fig. \ref{fig:pdf}). We attribute this portion of the distribution to the strongest forces close to the intruders which are generated as a direct result of the active dynamic intrusion, and would not be observed in systems under static equilibrium. We refer to this set of forces beyond the inflection point as "very strong forces". We calculated the number of very strong forces for different gap spacings to further explore the correlation between total force experienced by the intruders and the force chains within the particle domain. We found that the number of these forces, which typically are a part of the force chains, follow a pattern similar to the net work, $W$ done by the intruders (Fig. \ref{fig:pdf}B). This indicates that the presence of very strong forces between the intruders is likely responsible for the peak in force observed due to gap spacing.

During intrusions near a wall, force chains build from both the intruder surface and the wall, and eventually merged together as the intruder got closer to the wall \cite{Lim2017}. These observations indicate that the force chain topology should be influenced when two intruders are near each other.
The inset in figure \ref{fig:pdf}B show the normal forces between neighboring particles that are larger than the mean normal force, $<f_n>$, in the particle domain. These force chains show greater overlap between the two intruders near the peak force---suggesting greater interaction---that diminishes as the intruders are further separated. 


%% file: 4_conclusions.tex
\section{Conclusions}
Using a combination of laboratory experiment and DEM simulation, this study showed that the distance between neighboring intruders affects the total vertical force response to active intrusion into a granular substrate with a peak in the force response at an intruder gap spacing of  $s/d\sim3$. 

Initial experimental results suggested that this finding was robust to particle size and intrusion speed. Further exploration of these and other variables mostly supported this observation. Greater particle friction was associated with a larger force response, but $(s/d)_{peak}$ did not change with friction. In contrast, larger intruder width resulted in greater force generation and greater $(s/d)_{peak}$. The velocity profiles, $V_y$ and $V_z$, developed an inflection point between the intruders when $s/d>3$, due to changes in direction of granular flow at greater intruder distances. Examination of shear strain rate under the intruders showed overlapping high shear regions while $s/d\leq3$, which formed two separate regions at $s/d>3$. In comparison to other studies, Merceron \textit{et al}. has shown that the spacing of $s/d\sim3$ can alter the dynamics of particle rearrangements in a 2D granular packing and is independent of intruder size \cite{Merceron2018}. In a 3D system, two spheres separated by a distance of 3-4 particles experiences maximum repulsion forces relative to other separations \cite{dhiman2020origin}. Despite the differences in the problem setup, we all find that a separation of 3 particle diameters between intruders yield maximum differences in the parameter of study. 

Investigating the force chains between the granular particles during intrusion revealed the presence of a larger number of strong forces at separations corresponding to the peak force response. We also examined the role of intruder shape in force response, and it appeared to affect the extent of the production of very strong forces between the intruders, while accounting for the difference in force response at large separations. 

Taken together, these results indicate decreased interactions in granular flow and smaller force production for intruders at separation distances greater than $s/d\sim3$. Biologically, these findings could improve our understanding for how foot shape and interaction dynamics facilitate locomotion on granular substrates, and likewise, of the evolutionary processes leading to complex foot morphologies in animals \cite{Li2012,qian2015}.

%% file: 5_acknowledgments.tex
\section{Acknowledgments}

The authors would like to thank Shashank Agarwal and Ken Kamrin for helpful feedback. The research conducted in this study was supported by a grant from the National Science Foundation (IOS-1453106) to S.T.H., and Army Research Office to D.I.G. H.M.J. acknowledges support from the Army Research Office under under Grant Number W911NF-19-1-0245. This research also includes calculations carried out on Temple University's HPC resources and thus was supported in part by the National Science Foundation (CNS-1625061) and by the US Army Research Laboratory under contract number W911NF-16-2-0189.

%% file: 5_appendixA.tex
\section{}
\label{sec:appA}

\begin{figure}[ht]
\begin{center}
    \includegraphics[width=.4\textwidth, trim={0cm 10cm 22cm 0}]{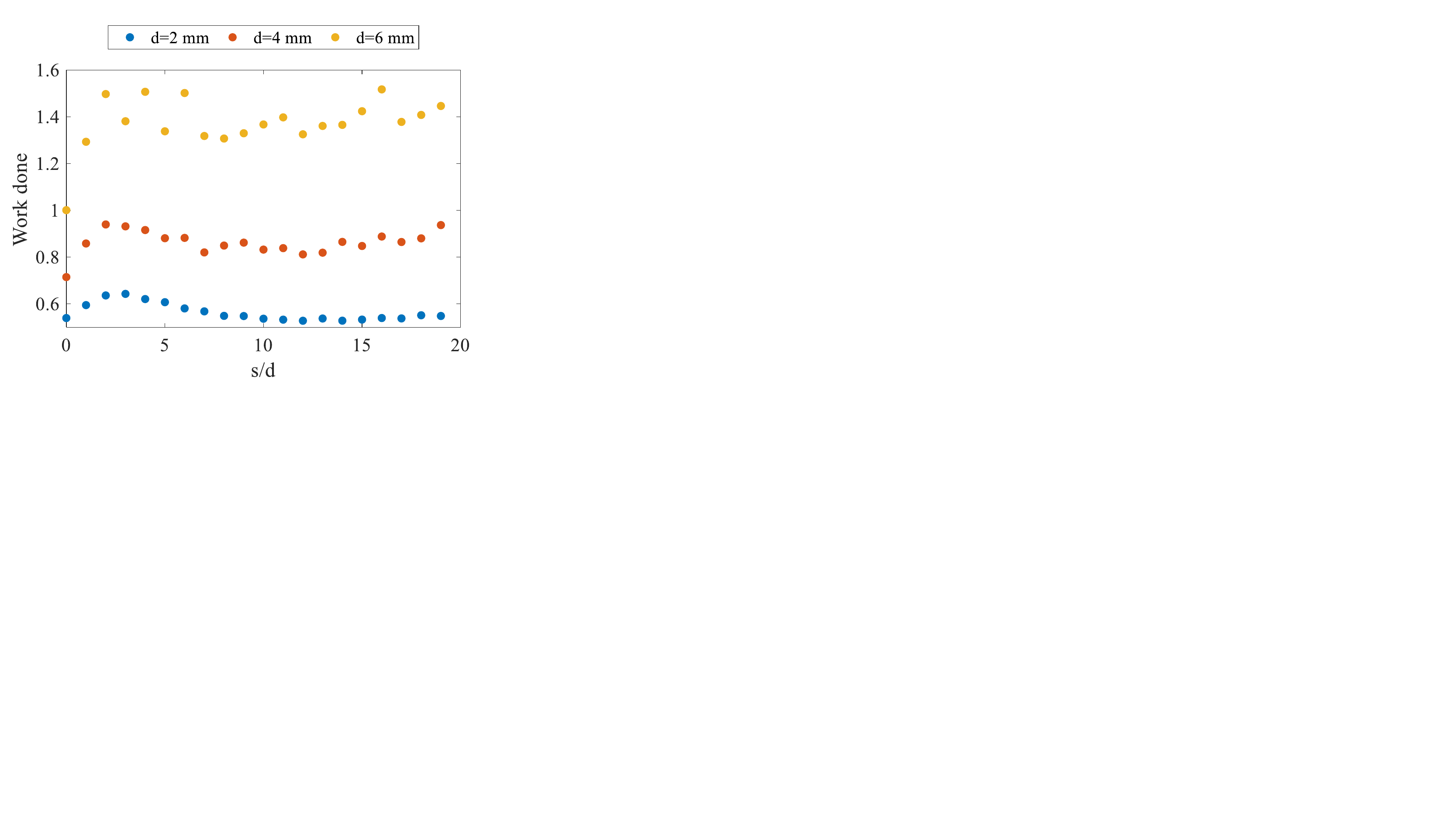}
    \caption{Work vs spacing for varying particle diameters. Intruder width, $D=5$ mm, and particle friction, $\mu= 0.5$, is held constant.}
    \label{fig:SI_particled}
\end{center}
\end{figure}

\begin{figure}[ht!]
\begin{center}
    \includegraphics[width=.4\textwidth, trim={0cm 10cm 22cm 0}]{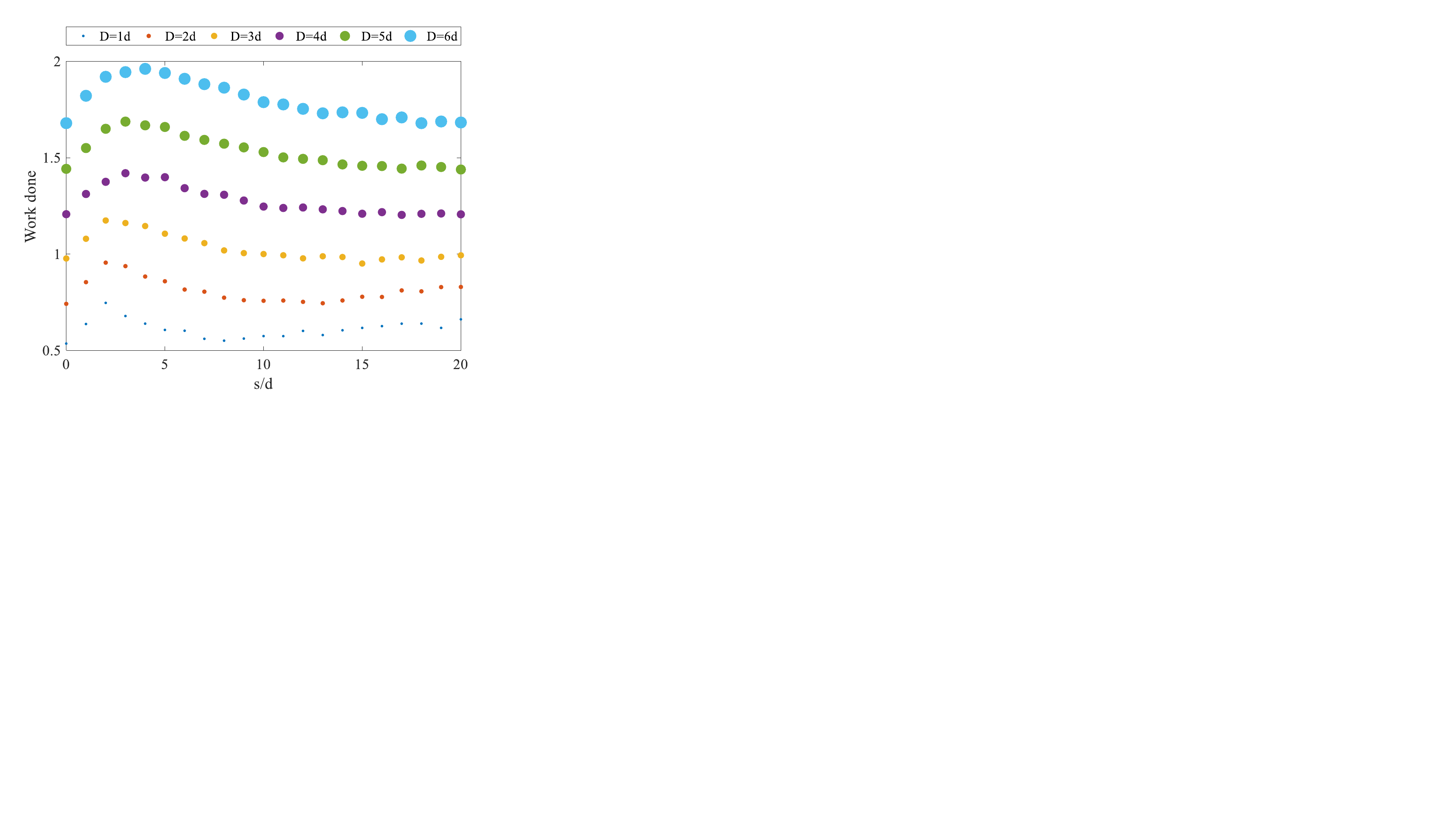}
    \caption{Work vs spacing for varying intruder widths. Particle diameter, $d=2$ mm, and particle friction, $\mu= 0.5$, is held constant.}
    \label{fig:SI_intruderD}
\end{center}
\end{figure}

\begin{figure}[ht]
\begin{center}
    \includegraphics[width=.4\textwidth, trim={0cm 10cm 22cm 0}]{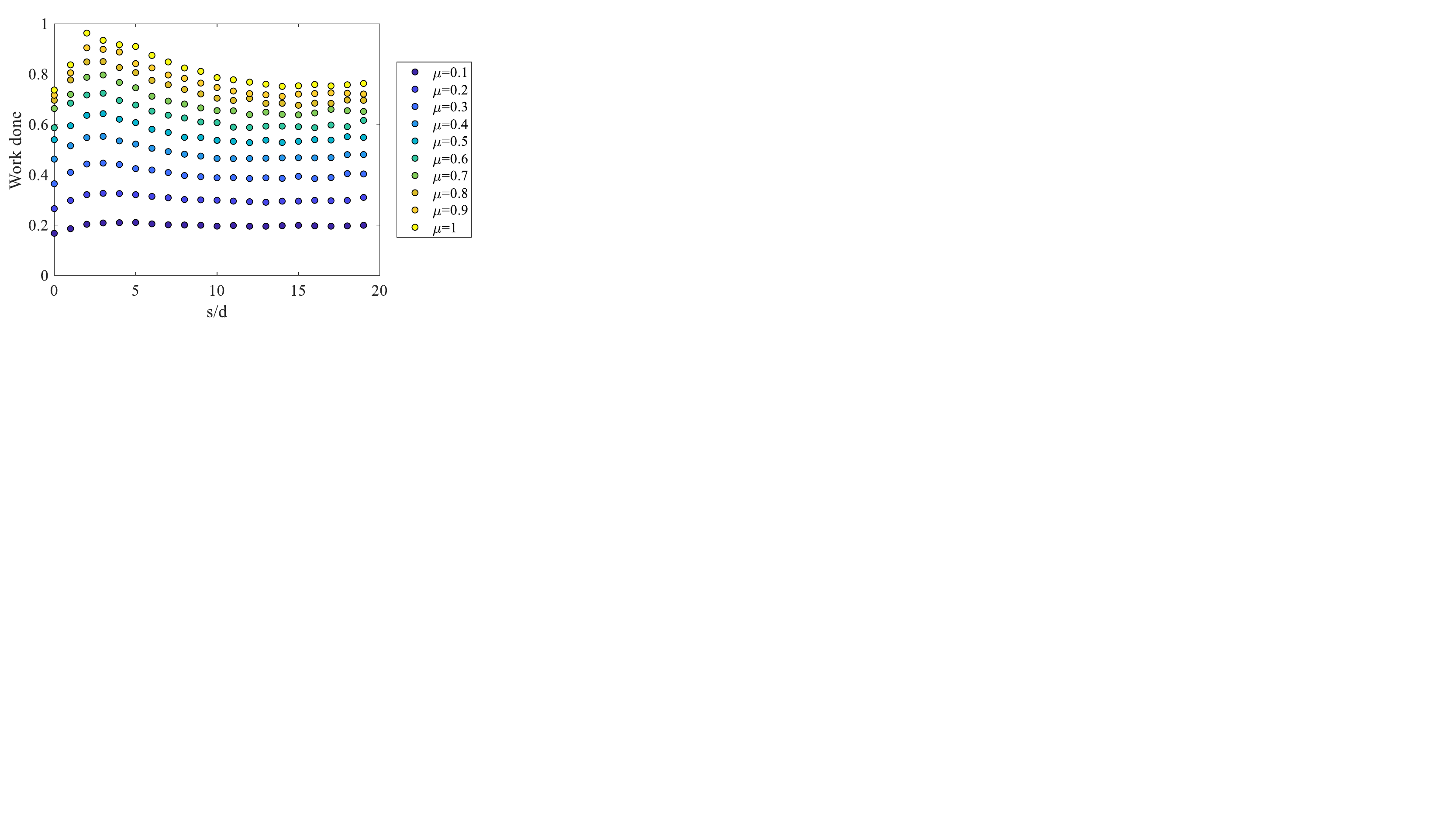}
    \caption{Work vs spacing for varying particle friction. Particle diameter, $d=2 mm$, and intruder width, $D=5$ mm, are held constant.}
    \label{fig:SI_friction}
\end{center}
\end{figure}

\begin{figure}[ht]
\begin{center}
    \includegraphics[width=.4\textwidth, trim={0cm 0cm 0cm 0}]{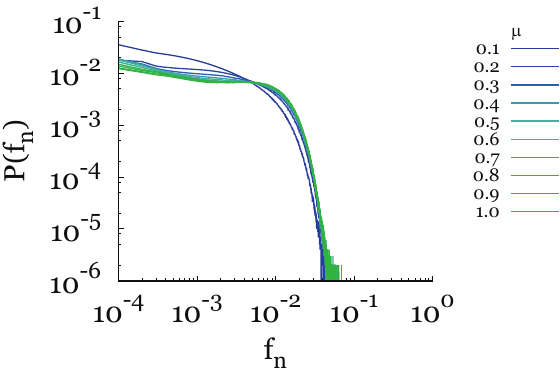}
    \caption{PDF of force chains with varying friction coefficient. Intruder spacing $s/d$ = 3, depth $z$ = 4cm, and other parameters are held constant.}
    \label{fig:SI_PDF}
\end{center}
\end{figure}

\begin{figure*}[ht!]
\begin{center}
    \includegraphics[width=.7\textwidth, trim={3.5cm 0 3.5cm 0}]{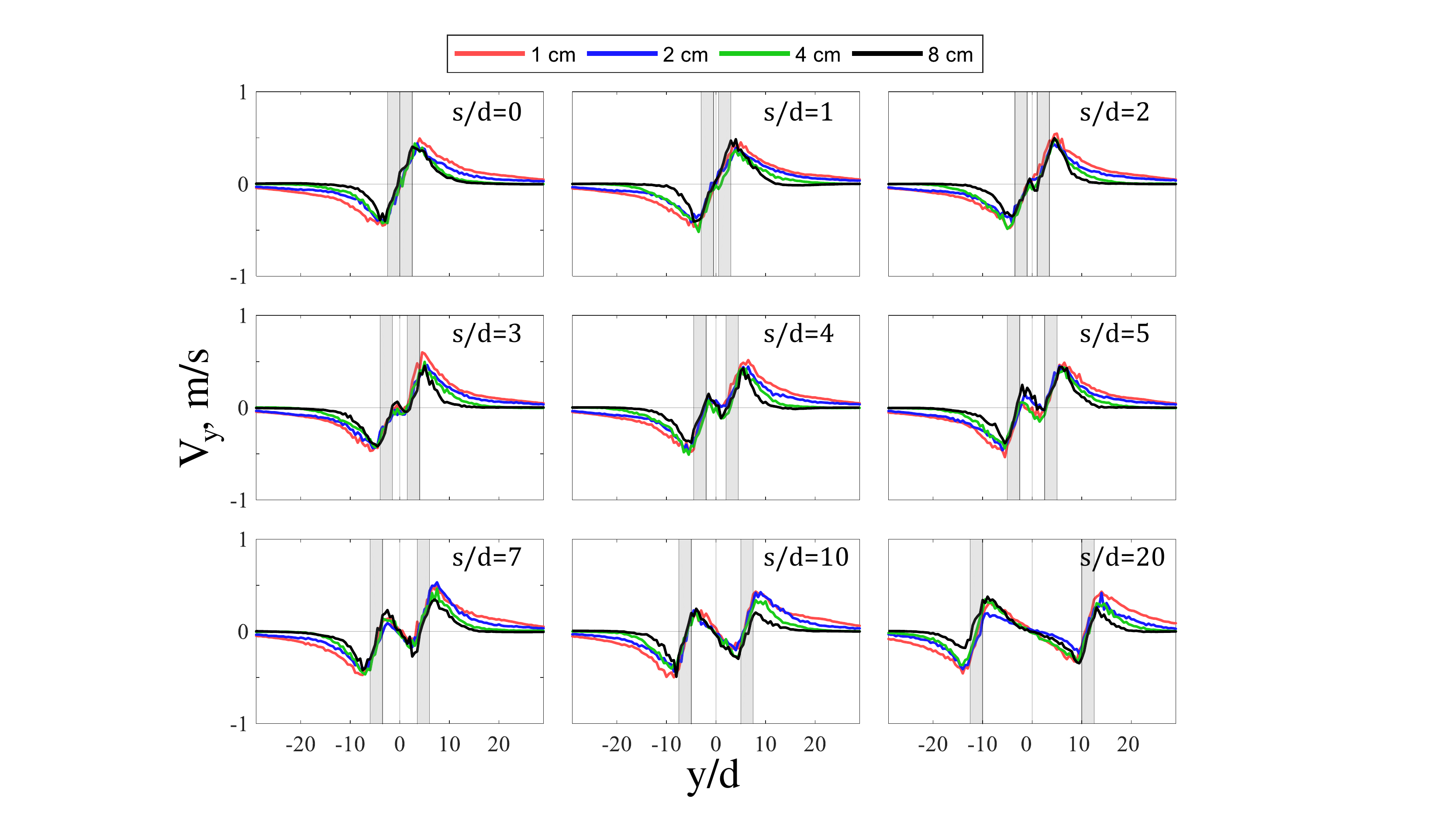}
    \caption{Y-direction velocity profile along y-direction for a variety of $s/d$ configurations. Vertical gray stripes indicate intruder boundaries. Depths of $z=$1, 2, 4, \& 8 cm are shown.}
    \label{fig:profile_vy}
\end{center}
\end{figure*}

\begin{figure*}[ht!]
\begin{center}
    \includegraphics[width=.7\textwidth, trim={3.5cm 0 3.5cm 0}]{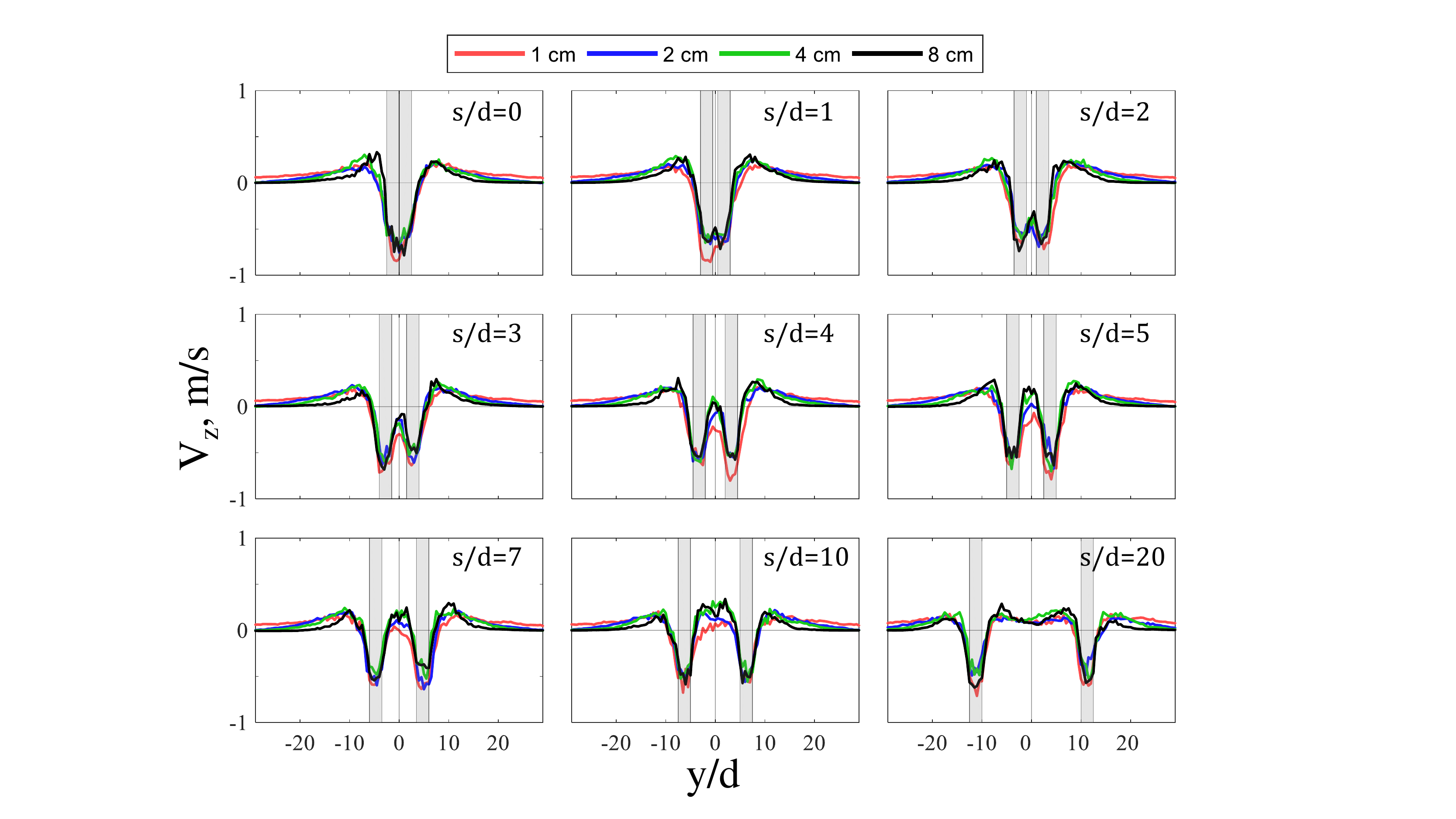}
    \caption{Z-direction velocity profile along y-direction for a variety of $s/d$ configurations. Vertical gray stripes indicate intruder boundaries. Depths of $z=$1, 2, 4, \& 8 cm are shown. }
    \label{fig:profile_vz}
\end{center}
\end{figure*}

%% file: 5_appendixB.tex
\section{}
\label{sec:appB}

\begin{align}
    k_n &= \frac{4}{3}Y^*\sqrt{R^*\delta_n} \\
    \gamma_n &= -2\frac{5}{6}\beta\sqrt{S_n m^*} \geq 0 \\
    k_t &= 8G^*\sqrt{R^*\delta_n} \\
    \gamma_t &= -2 \frac{5}{6}\beta \sqrt{S_t m^*} \geq 0
\end{align}

\begin{align}
    S_n &= 2Y^*\sqrt{R^*\delta_n} \\
    S_t &= 8G^*\sqrt{R^* \delta_n} \\
    \beta &= \frac{\log(e)}{\sqrt{\log^2(e)+\pi^2}} \\
    \frac{1}{Y^*} &= \frac{1-\nu_1^2}{Y_1} + \frac{1-\nu_2^2}{Y_2} \\
    \frac{1}{G^*} &= \frac{2(2-\nu_1)(1+\nu_1)}{Y_1} + \frac{2(2-\nu_2)(1+\nu_2)}{Y_2} \\
    \frac{1}{R^*} &= \frac{1}{R_1} + \frac{1}{R_2} \\
    \frac{1}{m^*} &= \frac{1}{m_1} + \frac{1}{m_2}
\end{align}

where $Y$ is the Young's modulus, $G$ is the shear modulus, $\nu$ is the Poisson's ratio, and $e$ is the coefficient of restitution. More details about the simulation method in LIGGGHTS can be found in \cite{kloss2012models}, and the contact-force models are described in articles by Di Renzo et al. \cite{di2004comparison, di2005improved}.